\documentclass[conference]{IEEEtran}
\usepackage{amsmath,amssymb}
\usepackage[dvips]{graphicx}
\usepackage{amsfonts}
\usepackage[mathscr]{eucal}
\usepackage{subfigure}
\usepackage{float}
\usepackage{latexsym}
\usepackage{amsthm}
\usepackage{exscale}
\usepackage[mathscr]{eucal}
\usepackage{bm}
\usepackage[dvipsnames]{color}
\usepackage{cases}
\usepackage{color}
\usepackage{epsfig}
\usepackage{stfloats}
\usepackage[ruled]{algorithm2e}
\usepackage[verbose,nospace,sort]{cite}
\usepackage{tabularx,ragged2e}
\usepackage{multirow}
\usepackage{multicol}
\usepackage{balance}
\usepackage{caption}
\usepackage{setspace}
\usepackage{multirow}
\usepackage{multicol,afterpage,wrapfig}
\usepackage{pifont}
\usepackage{array}

\theoremstyle{definition}

\usepackage{algpseudocode}

\graphicspath{{./figs/}}

\newtheorem{Prop}{Proposition}

\newenvironment{myproof} {{\it{Proof:}}}{\hfill$\square$}

\begin{document}

\title{Anchor-Assisted Intelligent Reflecting Surface Channel Estimation for Multiuser Communications}
\vspace{-0cm}
\author{\IEEEauthorblockN{Xinrong Guan\IEEEauthorrefmark{1}\IEEEauthorrefmark{2}, Qingqing Wu\IEEEauthorrefmark{3}, Rui Zhang\IEEEauthorrefmark{4}}
	\IEEEauthorblockA{\IEEEauthorrefmark{1}Communications Engineering College, Army Engineering University of PLA, Nanjing, 210007, China\\}	
	\IEEEauthorblockA{\IEEEauthorrefmark{2} Postdoctoral Station, Shenzhen Electric Appliance Company, Shenzhen, 518022, China\\}
	\IEEEauthorblockA{\IEEEauthorrefmark{3} Faculty of Science and Technology, University of Macau, Macau, 999078, China\\}
	\IEEEauthorblockA{\IEEEauthorrefmark{4} Department of Electrical and Computer Engineering, National University of Singapore, 117583, Singapore\\
	Email: geniusg2017@gmail.com, qingqingwu@um.edu.mo, elezhang@nus.edu.sg}\vspace{-7mm}
}

\maketitle

\begin{abstract} 
	Due to the passive nature of Intelligent Reflecting Surface (IRS), channel estimation is a fundamental challenge in IRS-aided wireless networks. Particularly, as the number of IRS reflecting elements and/or that of IRS-served users increase, the channel training overhead becomes excessively high. To tackle this challenge, we propose in this paper a new anchor-assisted two-phase channel estimation scheme, where two anchor nodes, namely A1 and A2, are deployed near the IRS for helping the base station (BS) to acquire the cascaded BS-IRS-user channels. Specifically, in the first phase, the partial channel state information (CSI), i.e., the element-wise channel gain square, of the BS-IRS link is obtained by estimating the BS-IRS-A1/A2 channels and the A1-IRS-A2 channel, separately. Then, in the second phase, by leveraging such partial knowledge of the BS-IRS channel that is common to all users, the individual cascaded BS-IRS-user channels are efficiently estimated. Simulation results demonstrate that the proposed anchor-assisted channel estimation scheme is able to achieve comparable mean-squared error (MSE) performance as compared to the conventional scheme, but with significantly reduced channel training time. 
\end{abstract}

\section{Introduction}
\vspace{-0mm}
Recently, intelligent reflecting surface (IRS) has emerged as a promising technology to achieve high spectral and energy efficiency for future wireless networks \cite{QQ}, \cite{Gong}. Specifically, IRS is a uniform planar array composed of a large number of low-cost, passive, and tunable reflecting elements. By adaptively varying the reflection coefficient of each element based on the user dynamic channels, IRS can achieve high beamforming and interference suppression gains  cost-effectively \cite{QQ_1}. As such, IRS has been studied recently in various wireless systems, including non-orthogonal multiple access (NOMA)\cite{zheng1}, \cite{ding}, simultaneous wireless information and power transfer (SWIPT) \cite{QQ_2}, \cite{pan}, secrecy communications \cite{guan}, \cite{shen}, and so on. 

To reap the performance gain of IRS, accurate channel state information (CSI) is required. However, the passive nature of IRS makes channel estimation fundamentally challenging in IRS-aided wireless networks. This is because without radio frequency (RF) chains, IRS can neither transmit nor receive pilot signals, thus the base station (BS)-IRS and IRS-user channels cannot be estimated separately. One alternative approach is to estimate the cascaded BS-IRS-user channel via element-wise on-off operation at each reflecting element \cite{yangyf}, \cite{mishra} or time-varying reflection patterns \cite{you}, \cite{zheng}. However, by applying the above methods to estimate the cascaded channels of multiple users consecutively, the required pilot overhead is the product of the number of IRS reflecting elements and that of users, which is prohibitively large for the case of large IRS serving a high density of users nearby (e.g., in a hot spot scenario).


\begin{figure}[t]
	\centering
	\includegraphics[width=3.3in]{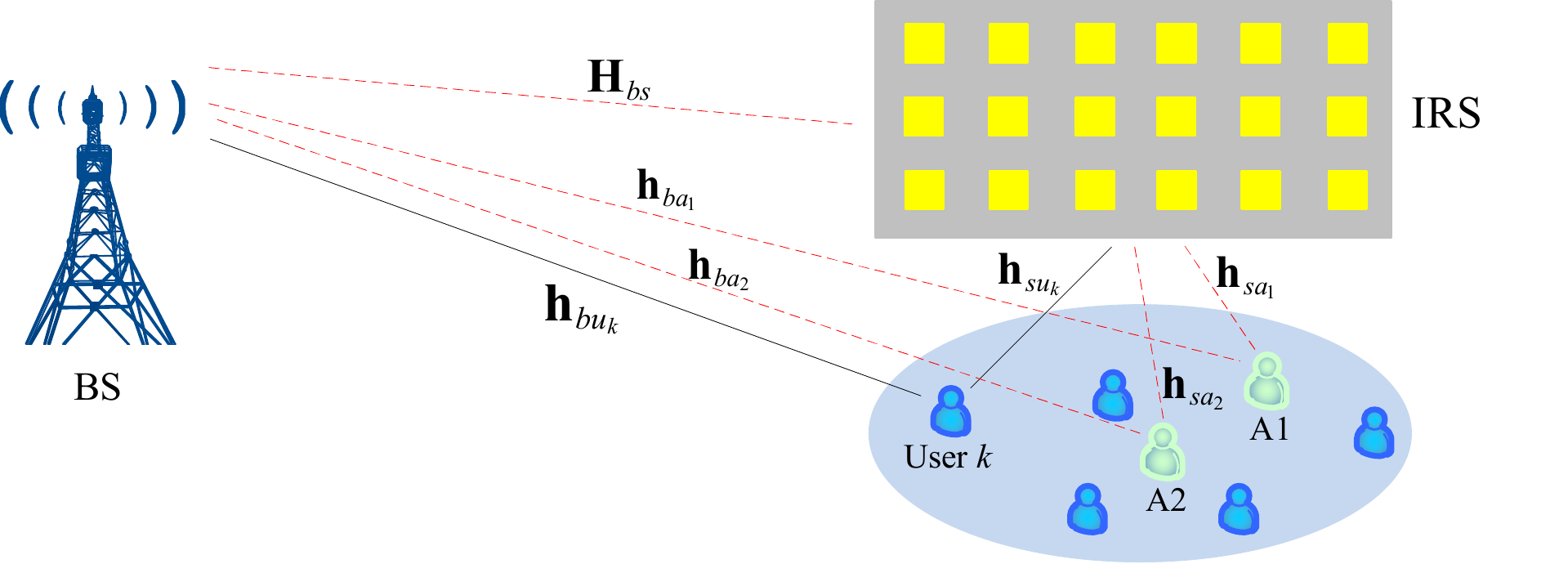}
	\caption{IRS-assisted multiuser MISO communication system.}
	\label{system_model}
	\vspace{-5mm}
\end{figure}

To tackle this problem, we propose in this paper a new anchor-assisted two-phase channel estimation scheme for IRS-aided multiuser communications, which can significantly reduce the channel training overhead by decoupling the estimation of cascaded channels and exploiting  multiple antennas at the BS. As shown in Fig. 1, two anchor nodes, namely A1 and A2, are deployed near the IRS to assist its channel estimation. In the first phase, the cascaded BS-IRS-A1/A2 and A1-IRS-A2 channels are estimated separately, based on which the partial CSI of the BS-IRS link, in terms of the square of each IRS element's channel with the BS (thus, with a +/- sign uncertainty), is obtained. In the second phase, with such partial CSI of the BS-IRS channel that is common to all users, the individual cascaded BS-IRS-user channels are efficiently estimated. Particularly, when the number of antennas at the BS ($M$) is no smaller than that ($N$) of the IRS reflecting elements, i.e. $M\!\ge\! N$, we show that the proposed scheme can estimate all users' cascaded channels in the second phase using only one pilot symbol. Besides, even for the case of $\!M<N\!$, the minimum training overhead (in terms of number of pilot symbols) is shown to be $\left\lceil\frac{NK}{M}\right\rceil$, where $K$ is the number of IRS-served users and $\left\lceil \cdot \right\rceil$ denotes the ceiling operation, which is significantly lower than $NK$ for the conventional schemes \cite{yangyf, mishra, you, zheng}. Moreover, in the case that the IRS-anchor channel is line-of-sight (LoS) by properly deploying the anchor, it is shown that the proposed scheme can be further simplified such that deploying only one anchor is sufficient. 


\begin{figure*}[t]
	\centering
	\includegraphics[width=7in]{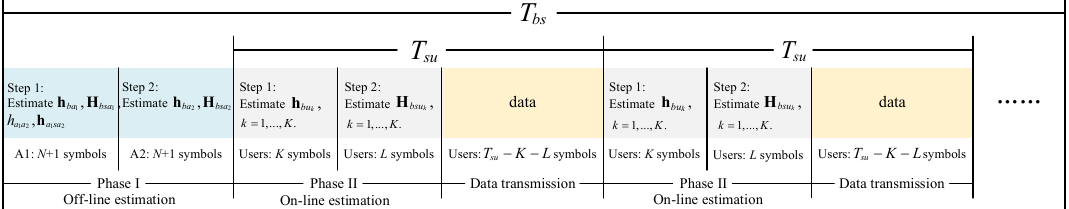}
	\caption{Chanel estimation and data transmission protocol.}\vspace{-5mm}
\end{figure*}

\section{System Model}
As shown in Fig. 1, we consider an IRS-assisted multiple-input-single-output (MISO) communication system, which consists of a BS, an IRS and $K$ users. The number of antennas at the BS and that of reflecting elements at the IRS are denoted by $M$ and $N$, respectively. The channels from the BS to IRS and User $k$ are denoted by ${\bf{H}}_{bs} \in{\mathbb{C}^{M \times N}}$ and ${\bf{h}}_{bu_{k}}\in{\mathbb{C}^{M \times 1}}$, respectively, while that from the IRS to User $k$ is denoted by ${\bf{h}}_{su_{k}}\in{\mathbb{C}^{N \times 1}}$. We assume that two single-antenna anchor nodes{\footnote{In practice, anchors can be idle user terminals and/or dedicated nodes such as adjacent IRS controllers.}} A1 and A2 are deployed near the IRS to assist in the channel estimation. The channels from the BS to A1 and A2 are denoted by ${\bf{h}}_{ba_1}\!\in\!{\mathbb{C}^{M \times 1}}$ and ${\bf{h}}_{ba_2}\!\in\!{\mathbb{C}^{M \times 1}}$, respectively, those from the IRS to A1 and A2 are denoted by ${\bf{h}}_{sa_1}\!\in\!{\mathbb{C}^{N \times 1}}$ and ${\bf{h}}_{sa_2}\!\in\!{\mathbb{C}^{N \times 1}}$, respectively, and that from A1 to A2 is denoted by ${{h}}_{a_1 a_2}$. As a result, the cascaded BS-IRS-A1/A2/User $k$ channels are denoted by ${\bf H}_{bsa_1}\!=\!{\bf H}_{bs}\text{diag}({\bf h}_{sa_1})$, ${\bf H}_{bsa_2}\!=\!{\bf H}_{bs}\text{diag}({\bf h}_{sa_2})$ and ${\bf H}_{bsu_k}={\bf H}_{bs}\text{diag}({\bf h}_{su_k})$, respectively, while the cascaded A1-IRS-A2 channel is denoted by ${\bf h}_{a_1sa_2}={\bf h}_{sa_2}^T\text{diag}({\bf h}_{sa_1})$. Moreover, the phase-shift matrix of the IRS at time slot $i$ is denoted by ${\mathbf{\Phi }}_i = \text{diag}\left( v_{1,i},v_{2,i},....,v_{N,i} \right)$, where $v_{n,i}$ is the reflection coefficient of the $n$-th IRS element at time slot $i$, $n=1,...,N$. Since IRS is a passive reflecting device, we assume that the channel reciprocity holds for each link between IRS and any other node. The quasi-static flat-fading channel model is assumed for all the links involved.

\vspace{-1.5mm}
\section{Anchor-assisted Channel Estimation}\vspace{-0.5mm}
The proposed channel estimation and data transmission protocol is shown in Fig. 2, where $T_{bs}$ and $T_{su}$ are the length of channel coherence block of ${\bf{H}}_{bs}$ and ${\bf{h}}_{su_{k}}$, $\forall k$, respectively, and $L=K$ for $M \ge N$ while $L=\left\lceil {\frac{KN}{{M}}} \right\rceil$ for $M<N$. In practice, $T_{bs}$ is usually much larger than $T_{su}$ due to the fixed locations of the BS and IRS once deployed, while users can move randomly near the IRS. The channel estimation consists of one off-line phase (Phase I) and multiple on-line phases (each termed Phase II). In Phase I, the cascaded BS-IRS-A1, BS-IRS-A2, A1-IRS-A2 channels are estimated separately, which requires $2(N+1)$ pilot symbols in total. In each Phase II, the BS estimates the cascaded BS-IRS-user channels with $K+L$ pilot symbols. For the estimation of ${\bf{h}}_{bsu_{k}}$ in Phase II, the efficiency is largely improved, especially when $M$ increases up to $M \ge N$. The detailed channel estimation scheme is elaborated as follows.

\vspace{-4mm}
\subsection{Phase I: Off-line Estimation of ${\bf{H}}_{bs}\odot {\bf{H}}_{bs}$}\vspace{-2mm}
\subsubsection{Step 1}A1 transmits pilot symbol $a_{1,i}$ with power $p$ at time slot $i$, then the received signals at the BS and A2 are respectively written as\vspace{-1mm}
\begin{equation}\small
{\bf{y}}_b^{(i)}=\sqrt{p}\left({\bf{h}}_{ba_1}+{\bf H}_{bs}{\bf{\Phi }}_i{\bf{h}}_{sa_1}\right)a_{1,i}+{\bf{z}}_{b}^{(i)},\label{y_b}\vspace{-1mm}
\end{equation}
\begin{equation}\small
{{{y}}_{2}^{(i)}} =\sqrt{p}\left( {{h}}_{a_1 a_2}+{\bf{h}}_{sa_2}^T{\bf{\Phi }}_i{\bf{h}}_{sa_1} \right)a_{1,i}  + {z_2^{(i)}}.\label{y_2}\vspace{-1mm}
\end{equation}
Let ${\bf{v}}_i=[v_{1,i},v_{2,i},....,v_{N,i}]^T$, ${\bf{\tilde{v}}}_i=[1, {\bf{v}}_i^T]^T$, ${\bf{\tilde{H}}}_{ba_1}=[{\bf{h}}_{ba_1}, {\bf{H}}_{bsa_1}]$, and ${\bf{\tilde{h}}}_{a_1a_2}=[h_{a_1a_2}, {\bf{h}}_{a_1sa_2}]$. Then, (\ref{y_b}) and (\ref{y_2}) are rewritten as\vspace{-1mm}
\begin{equation}\small
{\bf{y}}_b^{(i)}=\sqrt{p}{\bf{\tilde{H}}}_{ba_1} {{\bf{\tilde{v}}}_i} {a_{1,i}} +{\bf{z}}_{b}^{(i)},\vspace{-1mm}
\end{equation}
\begin{equation}\small
{{{y}}_{2}^{(i)}} =\sqrt{p} {\bf{\tilde{h}}}_{a_1a_2} {{\bf{\tilde{v}}}_i} {a_{1,i}}  + {z_2^{(i)}}.\vspace{-1mm}
\end{equation}
There are in total $(N+1)M$ unknowns in ${\bf{\tilde{H}}}_{ba_1}$ and the BS has $M$ observations at each time slot, thus A1 has to transmit $N+1$ pilot symbols at least for the BS to estimate ${\bf{\tilde{H}}}_{ba_1}$. Similarly, it can be shown that at least $N+1$ pilot symbols are needed for A2 to estimate ${\bf{\tilde{h}}}_{a_1a_2}$. By setting  $a_{1,i}=1$, $i=1,...,N+1$ and denoting ${{\bf{\tilde{V}}}}=[{{\bf{\tilde{v}}}_1},...,{{\bf{\tilde{v}}}_{N+1}}]$, the overall received signals at the BS and A2 during the $N+1$ time slots are given by
\begin{equation}\small
\label{Y}
{{\bf{Y}}_b}=[{\bf{y}}_b^{(1)},...,{\bf{y}}_b^{(N+1)}]=\sqrt{p}{\bf{\tilde{H}}}_{ba_1} {{\bf{\tilde{V}}}} +{\bf{Z}}_{b},
\end{equation}
\begin{equation} \small
{{\bf{y}}_{2}} =[{{{y}}_{2}^{(1)}},...,{{{y}}_{2}^{(N+1)}}]=\sqrt{p}{\bf{\tilde{h}}}_{a_1a_2} {{\bf{\tilde{V}}}} + {{\bf{z}}_2},
\end{equation}  
where ${\bf{Z}}_{b}=[{\bf{z}}_{b}^{(1)},...,{\bf{z}}_{b}^{(N+1)}]$ and ${{\bf{z}}_2}=[{z_2^{(1)}},...,{z_2^{(N+1)}}]$, respectively. By properly constructing ${{\bf{\tilde{V}}}}$ such that $\text{rank}({{\bf{\tilde{V}}}})=N+1$, ${\bf{\tilde{H}}}_{ba_1}$ and ${\bf{\tilde{h}}}_{a_1a_2}$ can be respectively estimated as\vspace{-1mm}
\begin{equation}\small
\label{H_ba1}
{\bf{\hat{H}}}_{ba_1}=[{\bf{\hat h}}_{ba_1}, {\bf{\hat H}}_{bsa_1}]=\frac{1}{\sqrt{p}}{{\bf{Y}}_b} {{\bf{\tilde{V}}}^{-1}},\vspace{-1mm}
\end{equation}
\begin{equation} \small
{\bf{\hat{h}}}_{a_1a_2} =[{\hat h}_{a_1a_2}, {\bf{{\hat h}}}_{a_1sa_2}]= \frac{1}{\sqrt{p}}{{\bf{y}}_{2}}{{\bf{\tilde{V}}}^{-1}}.
\vspace{-2mm}
\end{equation}
Practically, ${{\bf{\tilde{V}}}}$ can be constructed based on the discrete Fourier transform (DFT) matrix \cite{zheng}, i.e., \vspace{-1mm}
\begin{equation}\small
{{\bf{\tilde{V}}}}=\left[
\begin{aligned}
&1~~~&~1~~~~~~~                   &...~~&~1~~~~~                   \\
&1~~~&e^{-j\frac{2\pi}{N+1}}~~~   &...~~&e^{-j\frac{2\pi }{N+1}N}  \\
&1~~~&e^{-j\frac{2\pi}{N+1}2}~~~  &...~~&e^{-j\frac{2\pi }{N+1}2N} \\
&.   &~.~~~~~~~                   &...~~&~.~~~~~                   \\
&.   &~.~~~~~~~                   &...~~&~.~~~~~                   \\
&1~~~&e^{-j\frac{2\pi}{N+1}N}~~~  &...~~&e^{-j\frac{2\pi }{N+1}N^2} 
\end{aligned}
\right].\vspace{-1mm}
\end{equation}
In this case, ${{\bf{\tilde{V}}}^{-1}}$ can be efficiently computed as by ${{\bf{\tilde{V}}}^{-1}}=\frac{1}{N+1}{{\bf{\tilde{V}}}^H}$.

\setcounter{equation}{20}
\begin{figure*}[hb]
	\vspace{-3mm}
	\begin{equation}\small
	\label{y_case2}
	\left[ {\begin{array}{*{20}{c}}
		{{{\bf{y}}^{(1)}}}\\
		{{{\bf{y}}^{(2)}}}\\
		\begin{array}{l}
		.\\
		.
		\end{array}\\
		{{{\bf{y}}^{\left( N \right)}}}
		\end{array}} \right] =\sqrt{p} \underbrace {\left[ {\begin{array}{*{20}{c}}
			{{{\bf{H}}_{bs}}{{\bf{\Phi}} _1}{x_{1,1}}}&{{{\bf{H}}_{bs}}{{\bf{\Phi}} _1}{x_{2,1}}}&{{{\bf{H}}_{bs}}{{\bf{\Phi}} _1}{x_{3,1}}}&{...}&{{{\bf{H}}_{bs}}{{\bf{\Phi}} _1}{x_{M,1}}}\\
			{{{\bf{H}}_{bs}}{{\bf{\Phi}} _2}{x_{1,2}}}&{{{\bf{H}}_{bs}}{{\bf{\Phi}} _2}{x_{2,2}}}&{{{\bf{H}}_{bs}}{{\bf{\Phi}} _2}{x_{3,2}}}&{...}&{{{\bf{H}}_{bs}}{{\bf{\Phi}} _2}{x_{M,2}}}\\
			.&.&.&{...}&.\\
			.&.&{}&.&{}\\
			{{{\bf{H}}_{bs}}{{\bf{\Phi}} _N}{x_{1,N}}}&{{{\bf{H}}_{bs}}{{\bf{\Phi}} _N}{x_{2,N}}}&{{{\bf{H}}_{bs}}{{\bf{\Phi}} _N}{x_{3,N}}}&{...}&{{{\bf{H}}_{bs}}{{\bf{\Phi}} _N}{x_{M,N}}}
			\end{array}} \right]}_{{\bf{B}} \in {\mathbb{C}^{MN \times MN}}}\left[ {\begin{array}{*{20}{c}}
		{{{\bf{h}}_{s{u_1}}}}\\
		{{{\bf{h}}_{s{u_2}}}}\\
		\begin{array}{l}
		.\\
		.
		\end{array}\\
		{{{\bf{h}}_{s{u_M}}}}
		\end{array}} \right]\vspace{-3mm}
	\end{equation}
\end{figure*}
\setcounter{equation}{9}

\subsubsection{Step 2} With ${\bf{\tilde{H}}}_{ba_2}=[{\bf{h}}_{ba_2}, {\bf{H}}_{bsa_2}]$, A2 transmits at least $N+1$ pilot symbols so that the BS can estimate ${\bf{\tilde{H}}}_{ba_2}$ as 
${\bf{\hat{H}}}_{ba_2}=[{\bf{\hat h}}_{ba_2}, {\bf{\hat H}}_{bsa_2}]$, similarly as in Step 1 and thus the details are omitted. With ${\bf \hat h}_{a_1sa_2}$ fed back from A2, BS obtains the estimated BS-IRS-A1, BS-IRS-A2 and A1-IRS-A2 channels, which are given by
\begin{align}
{\bf{\hat H}}_{bsa_1}&={\bf{\hat H}}_{bs}{\rm{diag}}({\bf{\hat h}}_{sa_1}),\label{H_bra1}\\   
{\bf{\hat H}}_{bsa_2}&={\bf{\hat H}}_{bs}{\rm{diag}}({\bf{\hat h}}_{sa_2}),\label{H_bra2}\\   
{\bf{\hat h}}_{a_1sa_2}&={\bf{\hat h}}_{sa_2}^T{\rm{diag}}({\bf{\hat h}}_{sa_1}).\label{h_a1ra2}
\end{align}
Based on (\ref{H_bra1})-(\ref{h_a1ra2}), the BS computes\vspace{-1mm}
\begin{equation}
\label{HodotH}
{\bf{\hat H}}_{bs}\odot{\bf{\hat H}}_{bs}={\bf{\hat H}}_{bsa_1} \odot {\bf{\hat H}}_{bsa_2}({\rm{diag}}({\bf{\hat h}}_{a_1sa_2}))^{-1},
\end{equation}
where $\odot$ denotes the Hadamard product. 
By defining ${\bf{G}}={\bf{\hat H}}_{bsa_1} \odot {\bf{\hat H}}_{bsa_2}({\rm{diag}}({\bf{\hat h}}_{a_1sa_2})^{-1})$, ${\bf{\hat H}}_{bs}\odot{\bf{\hat H}}_{bs}$ is rewritten as
\begin{equation}\small
\label{H_br}
{\bf{\hat H}}_{bs}\odot{\bf{\hat H}}_{bs}
=\left[
\begin{aligned}
&{\bf{G}}_{11}&{\bf{G}}_{12}~~~&...&{\bf{G}}_{1N}~\\
&{\bf{G}}_{21}&{\bf{G}}_{22}~~~&...&{\bf{G}}_{2N}~\\
&~~.          &.~~~~~          &...&.~~~                \\
&~~.          &.~~~~~          &...&.~~~                \\
&{\bf{G}}_{M1}&{\bf{G}}_{M2}~~ &...&{\bf{G}}_{MN}
\end{aligned}
\right],
\end{equation}
where ${\bf G}_{mn}\!=\!{\bf \hat H}_{bs}(m,n)^2$. Letting $g_{mn}=\sqrt{|{\bf{G}}_{mn}|}e^{j \frac{\angle {\bf{G}}_{mn}}{2}}$, then each element in ${\bf{\hat H}}_{bs}$ can be obtained as\vspace{-1mm} 
\begin{equation}
\label{H_bs}
 {\bf{\hat H}}_{bs}(m,n)=\pm g_{mn}, \forall m, n, \vspace{-1mm}
\end{equation}
i.e., we recover each ${\bf{\hat H}}_{bs}(m,n)$ but with  a +/- sign uncertainty. However, we will show later that such partial CSI estimated is sufficient for resolving the cascaded channels ${\bf H}_{bsu_k}$'s of all users uniquely. 
\vspace{-2mm}
\subsection{Phase II: On-line Estimation of ${\bf{h}}_{bu_k}$ and  ${\bf{H}}_{bsu_k}$}\vspace{-1mm}
\subsubsection{Step 1}$K$ users sequentially transmit one pilot symbol while the BS estimates ${\bf{h}}_{bu_k}$, $\forall k$, respectively, where the details are omitted for brevity. 

\subsubsection{Step 2}Users transmit $\text{max}\left(K,\left\lceil  {\frac{KN}{M}} \right\rceil\right)$ pilot symbols while the BS estimates ${\bf{H}}_{bsu_k}$, $\forall k$, respectively. Denoting the pilot symbol transmitted from User $k$ at time slot $i$ by $x_{k,i}$, the received signal at the BS is given by\vspace{-2mm}
\begin{align}
{\bf{y}}_b^{(i)}=\sqrt{p}\sum_{k=1}^{K}\left({\bf{h}}_{bu_k}+{\bf H}_{bs}{\bf{\Phi }}_i{\bf{h}}_{su_k}\right)x_{k,i}+{\bf{z}}_{b}^{(i)}.\label{y_b_1}\vspace{-1mm}
\end{align}
Let ${\bf H}_{bs}{\bf{\Phi }}_i{\bf{h}}_{su_k}={\bf{ H}}_{bsu_k}{\bf{v }}_i$, where ${\bf H}_{bsu_k}={\bf H}_{bs}\text{diag}({\bf h}_{su_k})$ and ${\bf{\Phi }}_i=\text{diag}({\bf{v}}_i)$. Then, by removing the signal from the direct channel, ${\bf{y}}_b^{(i)}$ can be re-expressed as\vspace{-3mm}
\begin{equation}
{\bf{\bar y}}_b^{(i)}=\sqrt{p}\sum_{k=1}^{K}{\bf{\hat H}}_{bsu_k}{\bf{v }}_ix_{k,i}+{\bf{\bar z}}_{b}^{(i)},\label{y_b_2}\vspace{-3mm}
\end{equation}
where  \vspace{-3mm}
\begin{equation}
 {\bf{\bar z}}_{b}^{(i)}\!=\!{\bf{z}}_{b}^{(i)}+\sqrt{p}\sum_{k=1}^{K}\left({\bf{h}}_{bu_k}\!-\!{\bf{\hat h}}_{bu_k}\!+\!({\bf H}_{bsu_k}\!-\!{\bf{\hat H}}_{bsu_k}){\bf{v }}_i\right)x_{k,i}\notag\vspace{-2mm}
\end{equation} 
is the effective noise, including the channel estimation error in ${\bf{\hat h}}_{bu_{k}}$ and ${\bf{\hat H}}_{bsu_k}$. In conventional schemes \cite{yangyf, mishra, you, zheng}, $N$ pilot symbols are required to estimate each ${\bf{H}}_{bsu_k}$. However, in our proposed scheme, by leveraging the partial CSI of ${\bf{\hat H}}_{bs}$ obtained in Phase I, the efficiency in estimating ${\bf{H}}_{bsu_k}$ can be greatly improved. Specifically, we consider the following two cases.

{\bf{Case 1:}} $M\ge N$. In this case, users send pilot symbols consecutively for the BS to estimate ${\bf{h}}_{su_{k}}$, independently. Let ${\bf{\Phi }}_k={\bf{I}}$, $x_{k,k}=1$ and $x_{k_1,k}=0,k_1\ne k$. Then, ${\bf{\bar y}}_b^{(k)}$ is rewritten as
\begin{align}
\label{y_est}
{\bf{\bar y}}_b^{(k)}=\sqrt{p}{\bf{\hat H}}_{bs}{\bf{h}}_{su_k}+{\bf{\tilde z}}_{b}^{(k)},
\end{align}
where ${\bf{\tilde z}}_{b}^{(k)}={\bf{z}}_{b}^{(k)}+\sqrt{p}\left({\bf{h}}_{bu_k}-{\bf{\hat h}}_{bu_k}+({\bf H}_{bsu_k}-{\bf{\hat H}}_{bsu_k}){\bf{v }}_k\right)$ and ${{\bf v}_k}=[1,1,...,1]^T$. 

Since ${\bf{\hat H}}_{bs}(m,n)$, $\forall m,n$, has two possible values,  ${\bf{h}}_{su_k}$ cannot be uniquely estimated. However, we show that the estimation of the cascaded channel ${\bf H}_{bsu_k}={\bf H}_{bs}\text{diag}({\bf h}_{su_k})$ can be uniquely obtained by resorting to the following proposition. \vspace{-4mm}
\begin{Prop}
	By setting ${\bf{\hat H}}_{bs}$ as ${\bf{W}}$ where ${\bf{W}}(1,n)=g_{1n}$ and ${\bf{W}}(m,n)=\frac{{\bf{\hat H}}_{bsa_1}(m,n)}{{\bf{\hat H}}_{bsa_1}(1,n)}g_{1n}$, $\forall m,n$,  ${\bf{h}}_{su_k}$ can be estimated as
	\begin{align}\label{h_suk}
	{\bf{\hat h}}_{su_k}=\frac{1}{\sqrt{p}}({\bf{W}}^H{\bf{W}})^{-1}{\bf{W}}^H{\bf{\bar y}}_b^{(k)},k=1,...,K.
	\end{align}
	Then, ${\bf{H}}_{bsu_k}$ is uniquely estimated as 
	\begin{equation}
	{\bf{\hat H}}_{bsu_k}={\bf{\hat H}}_{bs}{\rm{diag}}({\bf{\hat h}}_{su_k}).
	\end{equation}
	
	\begin{myproof}
	Please see Appendix A. 
	\end{myproof}
\end{Prop}

Based on Proposition 1, it takes the BS $K$ pilot symbols to estimate all BS-IRS-user cascaded channels.

{\bf{Case 2:}} $M < N$. In this case, if all users transmit pilot symbols one by one, it takes the BS at least $\left\lceil {\frac{N}{M}} \right\rceil $ pilot symbols to estimate each ${\bf{h}}_{su_k}$ and thus the total overhead is $K\left\lceil {\frac{N}{M}} \right\rceil $. Thus, we propose a scheme based on orthogonal pilot symbols and orthogonal phase shifts for reducing the total overhead to $\left\lceil  {\frac{KN}{M}} \right\rceil$, which is detailed as follows.

{{Step (a): }}Divide the $K$ users into $L_1=\left\lceil  {\frac{K}{M}} \right\rceil$ groups, such that there are $M$ users in each of the first $L_1-1$ groups and $M_1$ ($M_1\le M$) users in the last group, i.e., $K=(L_1-1)M+M_1$.

{{Step (b): }}For each of the first $L_1-1$ groups, $N$ pilot symbols are used to estimate  the BS-IRS-user cascaded channels. Take the first group as an example. The received signals at the BS during $N$ time slots by removing those from the direct path and neglecting the noise can be written as (\ref{y_case2}) shown at the bottom of this page, 
where ${\bf{\Phi}}_i=\text{diag}({\bf v}_i)$ is the phase-shift matrix in the $i$th time slot and $x_{k,i}$ is the pilot symbol transmitted by the $k$th user in the $i$th time slot, $k \le M$, $i \le N$. 
As long as we design $\Phi_i$ and $x_{k,i}$ properly such that $\bf B$ given in (\ref{y_case2}) is full-rank, the IRS-user channels can be estimated as
\setcounter{equation}{21}
\begin{equation}\small
\label{h_ruk_est_2}
\left[ {\begin{array}{*{20}{c}}
	{{{\bf{\hat h}}_{s{u_1}}}}\\
	{{{\bf{\hat h}}_{s{u_2}}}}\\
	\begin{array}{l}
	.\\
	.
	\end{array}\\
	{{{\bf{\hat h}}_{s{u_M}}}}
	\end{array}} \right] = \frac{1}{\sqrt{p}}{\left( {{{\bf{B}}^H}{\bf{B}}} \right)^{ - 1}}{{\bf{B}}^H}\left[ {\begin{array}{*{20}{c}}
	{{{\bf{\bar y}}_b^{(1)}}}\\
	{{{\bf{\bar y}}_b^{(2)}}}\\
	\begin{array}{l}
	.\\
	.
	\end{array}\\
	{{{\bf{\bar y}}_b^{\left( N \right)}}}
	\end{array}} \right].
\end{equation}
For example, we can design the pilot symbols transmitted by the $M$ users during the $N$ time slots as
\begin{equation}\small
{\bf X}\!=\!\left[ {\begin{array}{*{20}{c}}
	1&1&1&{...}&1\\
	1&{{e^{ - j\theta }}}&{{e^{ - j2\theta }}}&{...}&{{e^{ - j\left( {M - 1} \right)\theta }}}\\
	.&.&.&{...}&.\\
	.&.&{}&.&{}\\
	1&{{e^{ - j\left( {N - 1} \right)\theta }}}&{{e^{ - j2\left( {N - 1} \right)\theta }}}&{...}&{{e^{ - j{\left( {M - 1} \right){\left( {N - 1} \right)}}\theta }}}
	\end{array}} \right],\notag
\end{equation}
where ${\bf X}_{ik}=x_{k,i}$ is the pilot symbol transmitted by User $k$ in the $i$-th time slot, while the phase shifts during the $N$ time slots are given by
\begin{equation}\small
{\bf V}=\left[ {\begin{array}{*{20}{c}}
	1&1&1&{...}&1\\
	1&{{e^{ - j\theta }}}&{{e^{ - j2\theta }}}&{...}&{{e^{ - j\left( {N - 1} \right)\theta }}}\\
	.&.&.&{...}&.\\
	.&.&{}&.&{}\\
	1&{{e^{ - j\left( {N - 1} \right)\theta }}}&{{e^{ - j2\left( {N - 1} \right)\theta }}}&{...}&{{e^{ - j{{\left( {N - 1} \right)^2}}\theta }}}
	\end{array}} \right],\notag
\end{equation}
where ${\bf V}_{ni}=v_{n,i}$ is the phase shift of the $n$-th reflecting element in the $i$-th time slot.

It should be noted that similar to {\bf{Case 1}}, the exact $\bf B$ is unknown due to the fact that we only have the estimation of ${\bf{H}}_{bs}\odot {\bf{H}}_{bs}$, instead of ${\bf{H}}_{bs}$. However, it is shown in Proposition 1 that the cascaded channels can be uniquely estimated as
\begin{equation*}
[{\bf{\hat H}}_{bsu_1},...,{\bf{\hat H}}_{bsu_M}]={\bf{B}}[{\rm{diag}}({\bf{\hat h}}_{su_1}),...,{\rm{diag}}({\bf{\hat h}}_{su_{M}})].
\end{equation*}

{{Step (c): }}For the last group, if $M_1=M$, the process is the same as that in Step (b) and another $N$ pilot symbols are needed. In this case, we have $L_1=\frac{K}{M}$ and thus the total overhead is $L_1N=\frac{KN}{M}$.

On the other hand, if $M_1<M$, another $N_1=\left\lceil  {\frac{M_1N}{M}} \right\rceil$ pilot symbols are required, $M_1<N_1<N$. Specifically, $\bf X$ and $\bf V$ are given by
\begin{equation*}\small
{\bf X}\!\!=\!\!\left[ {\begin{array}{*{20}{c}}
	1&1&1&{...}&1\\
	1&{{e^{ - j\theta }}}&{{e^{ - j2\theta }}}&{...}&{{e^{ - j\left( {M_1 - 1} \right)\theta }}}\\
	.&.&.&{...}&.\\
	.&.&{}&.&{}\\
	1&{{e^{ - j\left( {N_1 - 1} \right)\theta }}}&{{e^{ - j2\left( {N_1 - 1} \right)\theta }}}&{...}&{{e^{ - j{\left( {M_1 \!-\! 1} \right){\left( {N_1 \!-\! 1} \right)}}\theta }}}
	\end{array}} \right],
\end{equation*}
\begin{equation*}\small
{\bf V}\!=\!\left[ {\begin{array}{*{20}{c}}
	1&1&1&{...}&1\\
	1&{{e^{ - j\theta }}}&{{e^{ - j2\theta }}}&{...}&{{e^{ - j\left( {N_1 - 1} \right)\theta }}}\\
	.&.&.&{...}&.\\
	.&.&{}&.&{}\\
	1&{{e^{ - j\left( {N - 1} \right)\theta }}}&{{e^{ - j2\left( {N - 1} \right)\theta }}}&{...}&{{e^{ - j{{\left( {N_1 - 1} \right)\left( {N - 1} \right)}}\theta }}}
	\end{array}} \right],
\end{equation*}
and thus $\bf B\in {\mathbb{C}}^{MN_1\times M_1N}$ is re-written as
\begin{equation*}\small
{\bf B}\!\!=\!\!\!\left[ {\begin{array}{*{20}{c}}
	{{{\bf{H}}_{bs}}{{\bf{\Phi}} _1}{x_{1,1}}}&{{{\bf{H}}_{bs}}{{\bf{\Phi}} _1}{x_{2,1}}}&{...}&{{{\bf{H}}_{bs}}{{\bf{\Phi}} _1}{x_{M_1,1}}}\\
	{{{\bf{H}}_{bs}}{{\bf{\Phi}} _2}{x_{1,2}}}&{{{\bf{H}}_{bs}}{{\bf{\Phi}} _2}{x_{2,2}}}&{...}&{{{\bf{H}}_{bs}}{{\bf{\Phi}} _2}{x_{M_1,2}}}\\
	.&.&{...}&.\\
	.&.&.&{}\\
	{{{\bf{H}}_{bs}}{{\bf{\Phi}} _{N_1}}{x_{1,N_1}}}&{{{\bf{H}}_{bs}}{{\bf{\Phi}} _{N_1}}{x_{2,N_1}}}&{...}&{{{\bf{H}}_{bs}}{{\bf{\Phi}} _{N_1}}{x_{M_1,N_1}}}
	\end{array}} \right]\!.
\end{equation*}
It can be verified that $MN_1>M_1N$ and $\bf B$ is column full-rank, i.e., $\text{rank}({\bf B})=M_1N$, such that the left inverse of $\bf B$ exists and thus the IRS-user channels can be estimated similarly as (\ref{h_ruk_est_2}). The corresponding overhead is $(L_1-1)N+\left\lceil  {\frac{M_1N}{M}} \right\rceil=\left\lceil  {\frac{M(L_1-1)N+M_1N}{M}} \right\rceil=\left\lceil  {\frac{KN}{M}} \right\rceil$.

As a result, the overhead in estimating the cascaded BS-IRS-user channels for the case of $M<N$ is $\left\lceil  {\frac{KN}{M}} \right\rceil$.

\begin{figure}[t]
	\centering
	\includegraphics[width=3.1in]{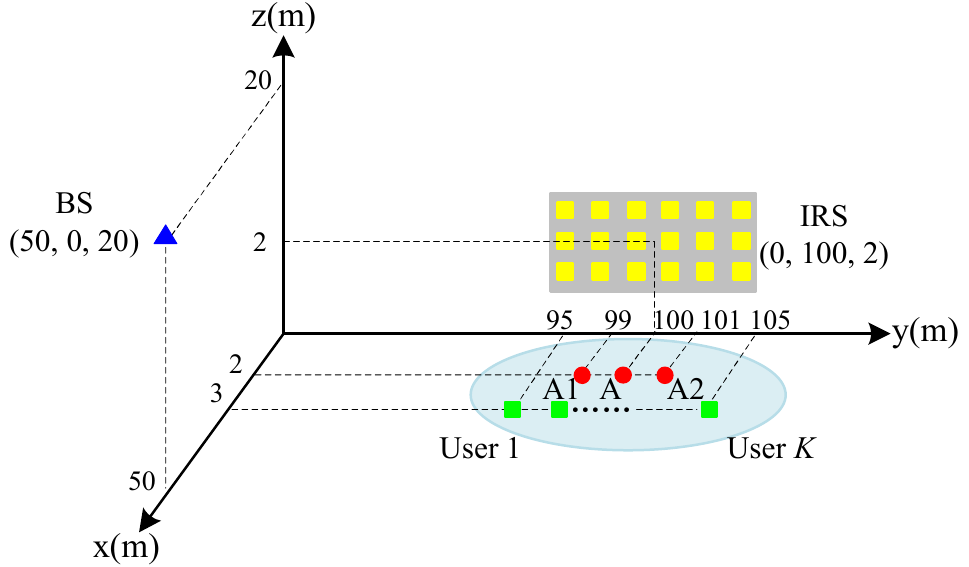}
	\caption{Simulation Setup.}
	\label{simulation_setup}
	\vspace{-5mm}
\end{figure}

\vspace{-2mm}
\subsection{Overall Training Overhead}\vspace{-1mm}
To summarize, the total training overhead of the proposed scheme is $2(N+1+K)$ for $M \ge N$ and $2(N+1)+K+\left\lceil\frac{KN}{M}\right\rceil$ for $M < N$, i.e. $2(N+1)+K+\text{max}(K,\left\lceil\frac{KN}{M}\right\rceil)$.

\section{Special Case: LoS IRS-Anchor Channel}
We assume that the IRS and/or the anchor node ``A'' can be properly deployed so that the IRS-A channel is LoS. In this case, ${\bf{h}}_{ra}$ is known a priori based on the knowledge of the positions of IRS and A, thus only one anchor node is sufficient for the proposed scheme and its training overhead can be further reduced.

\vspace{-2mm}
\subsection{Phase I: Off-line Estimation of ${\bf{H}}_{bs}$}\vspace{-1mm}
Let anchor A transmit $N+1$ pilot symbols while BS estimates ${\bf h}_{ba}$ and ${\bf H}_{bsa}={\bf H}_{bs}\text{diag}({\bf h}_{ra})$, respectively. Since ${\bf{h}}_{ra}$ is known, ${\bf{H}}_{bs}$ can be recovered from  ${\bf{\hat H}}_{bsa}$ as
\begin{equation}
{\bf{\hat H}}_{bs}={\bf{\hat H}}_{bsa} {\rm{diag}}({\bf{ h}}_{ra})^{-1}.\label{H_br_SC}
\vspace{-2mm}
\end{equation}

\subsection{Phase II: On-line Estimation of ${\bf{h}}_{bu_k}$ and ${\bf{H}}_{bsu_k}$}
\vspace{-1mm}

Users transmit pilot symbols while the BS estimates ${\bf{h}}_{bu_k}$ and ${\bf{H}}_{bsu_k}$, respectively. Specifically, we estimate the channel ${\bf{h}}_{su_k}$ first based on the estimation of ${\bf{\hat H}}_{bs}$ and then obtain  ${\bf{H}}_{bsu_k}={\bf \hat H}_{bs}\text{diag}({\bf \hat h}_{su_k})$, similar to Proposition 1. The only difference is that we have ${\bf{\hat H}}_{bs}$ in this case instead of ${\bf{\hat H}}_{bs} \odot {\bf{\hat H}}_{bs}$. 

Combining Phases I and II, the total training overhead of the proposed scheme in this special case is $N+1+2K$ for $M \ge N$ and $N+1+K+\left\lceil\frac{KN}{M}\right\rceil$ for $M < N$, i.e. $N+1+K+\text{max}(K,\left\lceil\frac{KN}{M}\right\rceil)$.


\begin{figure*}[t]
	\centering
	\subfigure[Training time versus $M$, with $K=20$.]{
		\includegraphics[width=2.04in]{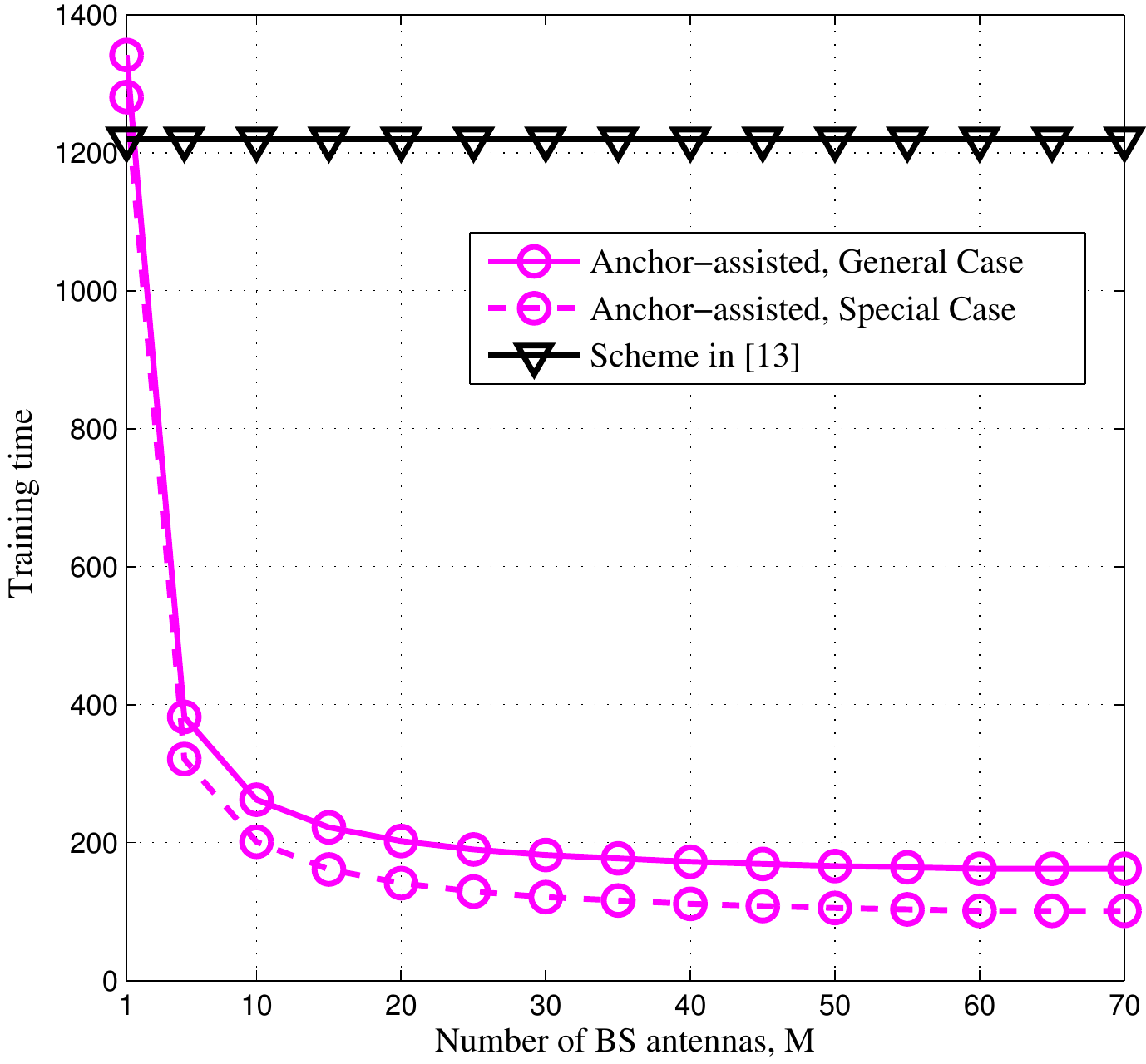}
		\label{fig:complexity_m}
	}	\hspace{-0mm}
	\subfigure[Normalized training time versus $K$.]{
		\includegraphics[width=2in]{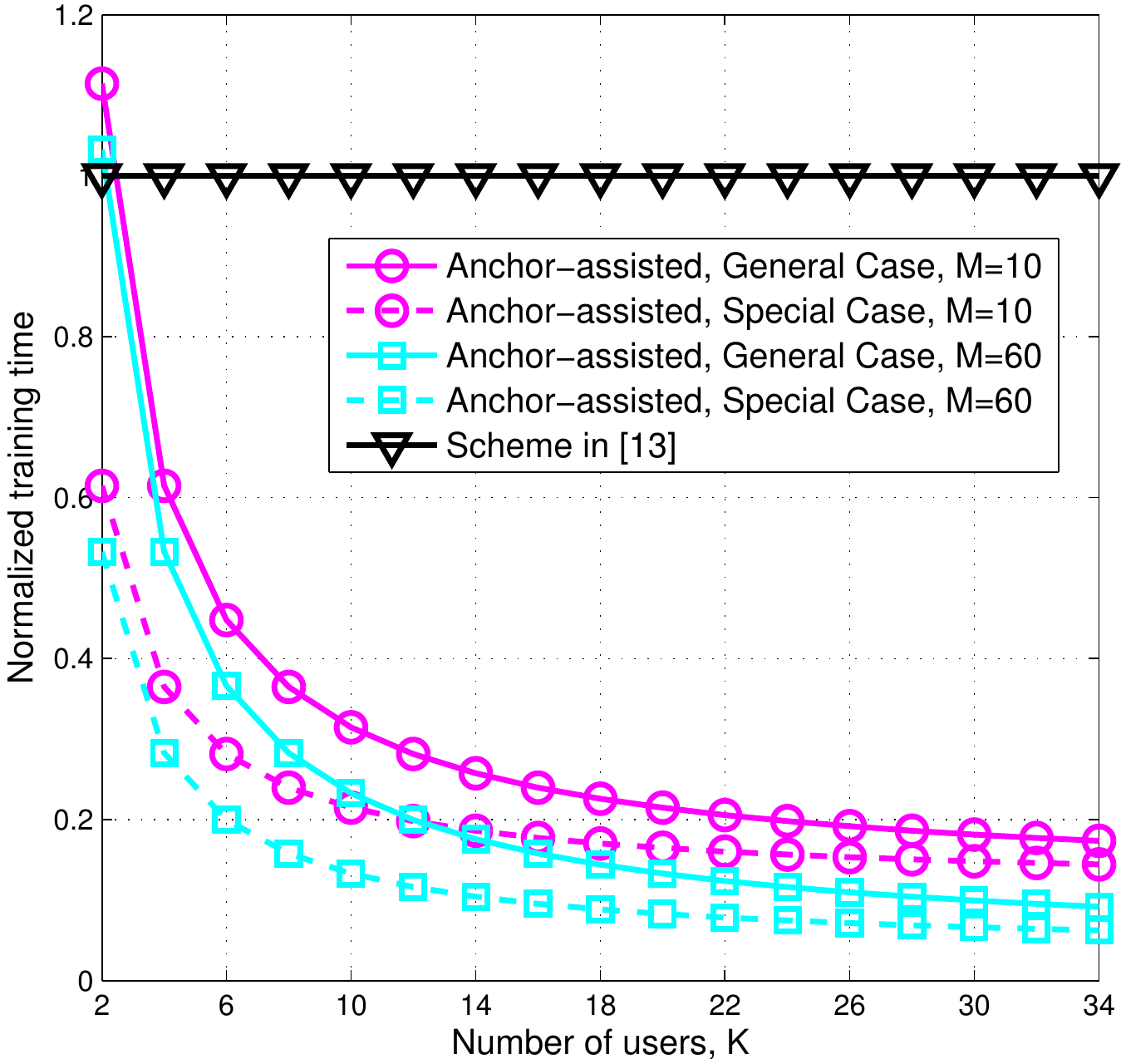}
		\label{fig:complexity_k}
	}\hspace{-0mm}
	\subfigure[MSE versus $p$, with $K=20$.]{
		\includegraphics[width=2in]{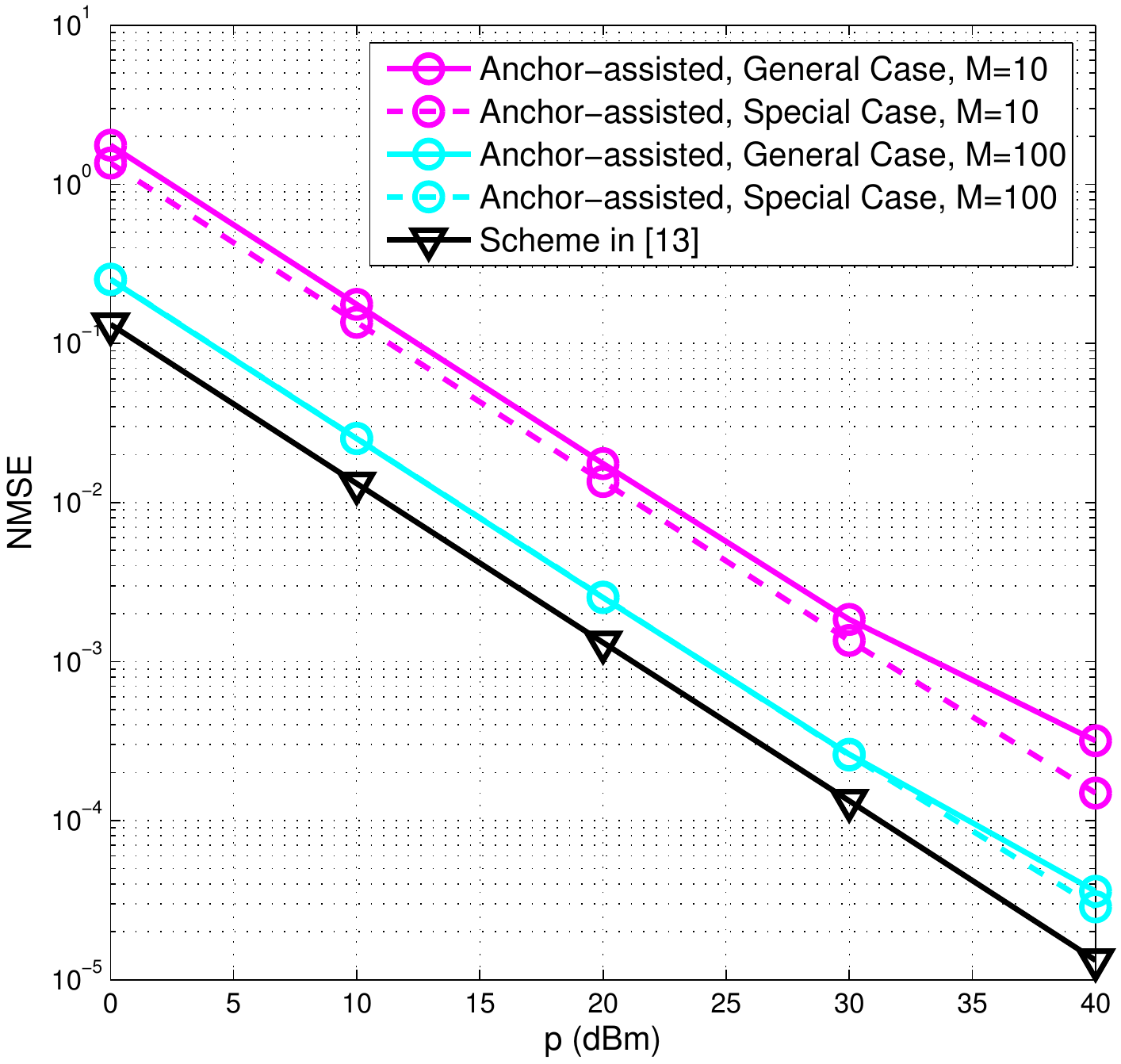}
		\label{fig:mse}
	}	\vspace{-3mm}
	\caption{Performance comparison between the proposed scheme and the benchmark scheme \cite{zheng}.}
	\label{performance}
	\vspace{-6mm}
\end{figure*}
\section{Numerical Results}
The simulation setup is shown in Fig. \ref{simulation_setup}. It is assumed that BS, IRS (the central point), A1, A2 and A are located at (50, 0, 20), (0, 100, 2),  (2, 99, 0), (2, 101, 0) and (2, 100, 0) in meter (m), respectively. We assume that the system operates on a carrier frequency of 750 MHz with the wavelength $\lambda_c=0.4$ m and the path loss at the reference distance $d_0\!=\!1$ m is given by $L_0=30$ dB. Suppose that the IRS is equipped with a uniform planar array with 6 rows and 10 columns, and the element spacing is $\Delta d\!=\! 3\lambda_c/8$; thus, we have $N=60$. The noise power is set as $\sigma_0^2\!=\!-$105 dBm. The channel from the BS to A1 is generated by ${{\mathbf{h}}_{ba_1}} \!=\! \sqrt {{L_0}d_{ba_1}^{ - {c_{ba_1}}}} {{\bf{g}}_{ba_1}}$, where $d_{ba_1}$ denotes the distance from the BS to A1 and ${{\bf{g}}_{ba_1}}$ is the small-scale fading component.
The same channel model is adopted for all other channels in general. Particularly, Rayleigh fading is assumed for the channels among the BS, IRS, A1, A2 and each User $k$ with the path loss exponents set as 3, whereas in the special channel case in Section IV, the channel between IRS and A is assumed to be LoS, with the path loss exponent set as 2. We also apply the scheme proposed in \cite{zheng} for each of the users to estimate their channels consecutively, which serves as the benchmark.  

Fig. \ref{fig:complexity_m} shows the required training time (in terms of number of pilot symbols) versus the number of antennas, $M$, at the BS. It is observed that as $M$ increases, the proposed scheme significantly reduce the training overhead as compared to that of the benchmark scheme (which is independent of $M$). This is because the proposed scheme exploits the multiple antennas at the BS for joint IRS channel estimation, whereas in the benchmark scheme the BS antennas estimate their associated channels independently in parallel. Note that when $M=1$, the training overhead of the proposed scheme is even larger than that of the benchmark scheme. This is because additional pilot symbols are transmitted by anchors in Phase I, while the training efficiency in Phase II is not improved since $\text{max}(K,\left\lceil\frac{NK}{M}\right\rceil)=NK$ in the case of $M=1$. Moreover, it is observed that the proposed scheme under the special case of LoS IRS-anchor channel is more efficient as compared to the general channel case. 

Fig. \ref{fig:complexity_k} shows the training time of the proposed scheme normalized by that of the benchmark scheme versus $K$, with $M=10$ and 60, respectively. One can observe that the performance gap between the two schemes becomes larger as $K$ increases. This is because to accommodate one more user, the additional pilot overhead required by the benchmark scheme is $N+1$, while that by the proposed scheme is $1+\text{max}(1,\left\lceil\frac{N}{M}\right\rceil)$. As a result, as $K$ increases, the pilot reduction by using the proposed scheme also increases. Also note that similar to Fig. \ref{fig:complexity_m}, when $K$ is very small, the proposed scheme is even worse than the benchmark scheme. This reason is that additional $2(N+1)$/$N+1$ pilot symbols are required in Phase I, regardless of $K$, while in the benchmark scheme, only $N+1$ pilot symbols are sufficient when $K=1$.

Fig. \ref{fig:mse} shows the normalized mean-squared error (MSE) of the estimations of ${\bf h}_{bu_k}$ and ${\bf H}_{bsu_k}$ versus the transmit power of pilot symbols in the on-line phase, with that of the off-line phase fixed as 40 dBm. It is observed that the MSE in the general channel case of the proposed scheme is highest, while that of the benchmark scheme is lowest. The reason is that although the proposed scheme significantly reduces the training overhead (see Figs. \ref{fig:complexity_m}  and \ref{fig:complexity_k}), the estimation error in ${\bf \hat H}_{bsu_k}$ depends on both ${\bf \hat H}_{bs}$ and ${\bf \hat h}_{su_k}$. Specifically, in the general channel case, the error in ${\bf \hat H}_{bs} \odot {\bf \hat H}_{bs}$ comes from ${\bf \hat H}_{bsa_1}$, ${\bf \hat H}_{bsa_2}$ and ${\bf \hat h}_{a_1sa_2}$, while in the special channel case the error in ${\bf \hat H}_{bs}$ comes from ${\bf \hat H}_{bsa}$. In contrast, for the benchmark scheme, ${\bf H}_{bsu_k}$ is estimated directly, which is thus less susceptible to noise/error. However, one can observe that the MSE of the proposed scheme is substantially reduced by increasing $M$, and becomes even comparable when $M=100$.

%
\vspace{-2mm}

\section{Conclusion}
\vspace{-2mm}

In this paper, we propose a new anchor-assisted channel estimation scheme for IRS-aided multiuser communications. By exploiting the fact that all BS-IRS-user cascaded channels share the same BS-IRS channel, the proposed scheme first estimates this common channel with only sign ambiguity via the anchor-assisted training. Then we show that the estimation of each cascaded BS-IRS-user channel is simplified to estimating each IRS-user channel with the number of unknowns significantly reduced from $MN$ to $N$, and the sign ambiguity in the estimated common channel does not affect the uniqueness of the recovered cascaded channels. Moreover, by exploring multi-antennas at the BS, the training overhead in estimating all users' cascaded channels is reduced from $NK$ to $\text{max}(K,\left\lceil\frac{NK}{M}\right\rceil)$. Numerical results validate the effectiveness of the proposed scheme, especially when $M$ and/or $K$ is large. Considering the trend towards massive antenna arrays at the BS and massive connectivity with machine-type communications, our proposed scheme has the great potential of significantly improving the channel estimation efficiency in future IRS-aided wireless systems.



\vspace{0mm}
\begin{appendices}
\section{}	
\vspace{-0mm}
First, we show the following lemma.

{\bf{Lemma 1:}} Given ${\bf{\hat H}}_{bsa_1}$, ${\bf{\hat H}}_{bs}(m,n)$ can be estimated as ${\bf{\hat H}}_{bs}(m,n)=\alpha_{mn}{\bf{\hat H}}_{bs}(1,n)$, where $\alpha_{mn}=\frac{{\bf{\hat H}}_{bsa_1}(m,n)}{{\bf{\hat H}}_{bsa_1}(1,n)}$, $\forall m$.

\begin{myproof}
	Because ${\bf{ H}}_{bsa_1}(1,n)={{\bf{ H}}_{bs}(1,n)}{{\bf{ h}}_{sa_1}(n)}$ and ${\bf{ H}}_{bsa_1}(m,n)={{\bf{ H}}_{bs}(m,n)}{{\bf{ h}}_{sa_1}(n)}$, we have ${\bf{ H}}_{bs}(m,n)=\frac{{\bf{ H}}_{bsa_1}(m,n)}{{\bf{ H}}_{bsa_1}(1,n)}{\bf{ H}}_{bs}(1,n)$, which thus completes the proof.  
\end{myproof}

Lemma 1 reveals that for given ${\bf{\hat H}}_{bsa_1}$, there are $N$ rather than $MN$ unknowns in  ${\bf{\hat H}}_{bs}$ and it can be rewritten as
\begin{equation*}\small
\label{H_bs_2}
{\bf{\hat H}}_{bs}\!=\!\left[
\begin{aligned}
&~~{{\bf{ H}}_{bs}(1,1)}         &{{\bf{ H}}_{bs}(1,2)}~~         &~~...&{{\bf{ H}}_{bs}(1,N)}~~\\
&\alpha_{21}{{\bf{ H}}_{bs}(1,1)}&\alpha_{22}{{\bf{ H}}_{bs}(1,2)}&~~...&\alpha_{2N}{{\bf{ H}}_{bs}(1,N)}\\
&~~~~~~~.            &.~~~~~~~~           &~~...&.~~~~~~~~                \\
&~~~~~~~.            &.~~~~~~~~           &~~...&.~~~~~~~~               \\
&\alpha_{M1}{{\bf{ H}}_{bs}(1,1)}&\alpha_{M2}{{\bf{ H}}_{bs}(1,2)}&~~...&\alpha_{MN}{{\bf{ H}}_{bs}(1,N)}
\end{aligned}
\right].
\end{equation*}
Meanwhile, we can construct a candidate of ${\bf H}_{bs}$ as ${\bf W}$ in Proposition 1, given by
\begin{equation}\small
\label{W}
{\bf{W}}=\left[
	\begin{aligned}
	&~~g_{11}         &g_{12}~~         &~~...&g_{1N}~~\\
	&\alpha_{21}g_{11}&\alpha_{22}g_{12}&~~...&\alpha_{2N}g_{1N}\\
	&~~~~.            &.~~~~~           &~~...&.~~~~~                \\
	&~~~~.            &.~~~~~           &~~...&.~~~~~                \\
	&\alpha_{M1}g_{11}&\alpha_{M2}g_{12}&~~...&\alpha_{MN}g_{1N}
	\end{aligned}
	\right].
\end{equation}

Next, we consider the following two cases.

{\bf Case 1:} ${\bf H}_{bs}={\bf W}$. In this case, we have ${{\bf{ H}}_{bs}(1,n)}=g_{1n}, \forall n$. Omitting the noise, we can express (\ref{y_est}) as \vspace{-1mm}
\begin{equation}
\label{case_a}
{\bf{\bar y}}_b^{(k)}=\sqrt{p}{\bf{W}}{\bf{h}}_{su_k} .\vspace{-1mm}
\end{equation}
Assuming that ${\bf W}$ is full-rank, ${\bf{h}}_{su_k} $ can be estimated as \vspace{-1mm}
\begin{equation}
\label{hat_h_suk}
{\bf{\hat h}}_{su_k}=\frac{1}{\sqrt{p}}({\bf{ W}}^H{\bf{W}})^{-1}{\bf{W}}^H{\bf{\bar y}}_b^{(k)},k=1,...,K.
\end{equation}
Then, the cascaded BS-IRS-user channel is estimated as 
\begin{equation}
\label{hat_H_bsuk_1}
{\bf{\hat H}}_{bsu_k}={\bf{ W}}\text{diag}({\bf{\hat h}}_{su_k})={\bf{\hat H}}_{bs}\text{diag}({\bf{ h}}_{su_k}). 
\end{equation}

{\bf Case 2:} ${\bf H}_{bs} \ne {\bf W}$. Referring to (\ref{H_bs}), there must exist at least an $n$ such that ${{\bf{ H}}_{bs}(1,n)}=-g_{1n}$. For illustration purpose, we assume that ${{\bf{ H}}_{bs}(1,1)}=-g_{11}$ and ${{\bf{ H}}_{bs}(1,n)}=g_{1n}$ for $\forall n \ne 1$, while for all other possible ${{\bf{ H}}_{bs}}$'s, the result can be similarly proved. Then we have 
\begin{equation}\small
\label{H_bs_3}
{\bf{\hat H}}_{bs}=\left[
\begin{aligned}
	&~-g_{11}         &g_{12}~~         &~~...&g_{1N}~~\\
	&-\alpha_{21}g_{11}&\alpha_{22}g_{12}&~~...&\alpha_{2N}g_{1N}\\
	&~~~~~~.            &.~~~~           &~~...&.~~~~~                \\
	&~~~~~~.            &.~~~~           &~~...&.~~~~~                \\
	&-\alpha_{M1}g_{11}&\alpha_{M2}g_{12}&~~...&\alpha_{MN}g_{1N}
\end{aligned}
\right].
\end{equation}
By omitting the noise, (\ref{y_est}) can be written as
\begin{equation}
  {\bf{\bar y}}_b^{(k)}=\sqrt{p}{\bf{\hat H}}_{bs}{\bf{h}}_{su_k}.
\end{equation}
Accordingly, by using ${\bf W}$, ${\bf h}_{su_k}$ is estimated as
\begin{equation}
{\bf{\hat h}}_{su_k}^{(2)}=({\bf{ W}}^H{\bf{W}})^{-1}{\bf{W}}^H{\bf{\hat H}}_{bs}{\bf{h}}_{su_k},k=1,...,K.\label{hat_h_suk_2}
\end{equation}
Comparing ${\bf{\hat H}}_{bs}$ in (\ref{H_bs_3}) with ${\bf{W}}$, we observe that the only difference lies in the sign of the elements in the first column and thus the following equality holds
\begin{equation}\vspace{-1mm}
{\bf{\hat H}}_{bs}={\bf{ W}}\text{diag}([-1,1,1,...,1]).\label{H_bs_4}
\end{equation}	
Since we have\vspace{-0mm}
\begin{equation}\label{H_bs_5}
[({\bf{\hat W}}^H{\bf{ W}})^{-1}{\bf{ W}}^H]{\bf{W}}={\bf{I}}_N,\vspace{-0mm}
\end{equation}
substituting (\ref{H_bs_4}) into (\ref{H_bs_5}) yields \vspace{-0mm}
\begin{equation}\label{W_H_bs}
[({\bf{ W}}^H{\bf{ W}})^{-1}{\bf{W}}_{2}^H]{\bf{\hat H}}_{bs}=\text{diag}([-1,1,1,...,1]).\vspace{-0mm}
\end{equation}
Based on (\ref{hat_h_suk_2}) and (\ref{W_H_bs}), the estimation of ${\bf{ h}}_{su_k}$ obtained by using ${\bf W}$  can be written as\vspace{-0mm}
\begin{equation}
	{\bf{\hat h}}_{su_k}^{(2)}=\text{diag}([-1,1,1,...,1]){\bf{ h}}_{su_k},k=1,...,K.\vspace{-0mm}
\end{equation}
Though ${\bf{\hat h}}_{su_k}^{(2)}$ is not an exact estimation, the error only occurs in the sign of the first element of ${\bf{ h}}_{su_k}$. Finally, the cascaded BS-IRS-user channel is recovered by\vspace{-0mm}
\begin{equation}
	{\bf{\hat H}}_{bsu_k}={\bf{W}}\text{diag}({\bf{\hat h}}_{su_k}^{(2)})={\bf{ \hat H}}_{bs}\text{diag}({\bf{h}}_{su_k}),k=1,...,K,\notag\vspace{-1mm}
\end{equation}
which is the same as (\ref{hat_H_bsuk_1}).

Based on Cases 1 and 2, it is concluded that using ${\bf W}$ constructed in Proposition 1 to estimate ${\bf{ H}}_{bsu_k}$ is always sufficient, regardless of whether ${\bf W}$ is exactly the same as ${\bf \hat H}_{bs}$, which thus completes the proof of Proposition 1.

\end{appendices}	
\ifCLASSOPTIONcaptionsoff
\newp\\a\apage
\fi

\end{document}